\documentclass[letter]{aa} 

\usepackage{graphicx}
\usepackage{txfonts}
\usepackage[allcolors=blue]{hyperref}
\begin{document}

   \title{Flybys in debris disk systems with {\it  Gaia} eDR3}

    \author{L. Bertini\inst{\ref{inst1}} 
            \and
            V. Roccatagliata\inst{\ref{inst3}, \ref{inst1}, \ref{inst4}}
            \fnmsep\thanks{veronica.roccatagliata@inaf.it} 
            \and
            M. Kim\inst{\ref{inst2}, \ref{inst5}, \ref{inst6}}}

    \institute{Department of Physics "E. Fermi" University of Pisa, Largo Bruno Pontecorvo 3, 56127 Pisa, Italy \label{inst1}     \and 
    INAF Osservatorio Astrofisico di Arcetri, Largo E. Fermi 5, 50125 Florence, Italy \label{inst3} \and 
    INFN, Section of Pisa, Largo Bruno Pontecorvo 3, 56127 Pisa, Italy \label{inst4} 
    \and 
    Department of Physics, University of Warwick, Gibbet Hill Road, Coventry CV4 7AL, UK \label{inst2}
    \and
    Centre for Exoplanets and Habitability, University of Warwick, Gibbet Hill Road, Coventry CV4 7AL, UK \label{inst5}
    \and
    Space Research Institute of the Austrian Academy of Sciences, Schmiedlstrasse 6, 8042 Graz, Austria \label{inst6}}

   \date{Received November 9, 2022; accepted January 19, 2023}

 
  \abstract
   {Debris disks represent the last phase of the evolution of protoplanetary disks around young stellar objects where planetary systems had most likely already been formed. Resolved systems show peculiar structures, such as asymmetries or spirals, which may be associated with either  the presence of a low-mass companion or dynamical interactions with a perturber during a flyby event.}
   {We aim to observationally and statistically constrain the influence of flybys in the formation and evolution of debris disks.}
   {We compiled a sample of 254 debris disks 
   with ages between 2 Myr and 8 Gyr that are either part of an association or isolated, drawing the binary and planetary companions of the systems mainly from the literature. Using the {\it Gaia} eDR3 astrometric data and radial velocities of our sample, as well as all the sources in a specific region of the sky, we reconstructed the relative linear  motions in the last 5 Myr and made predictions  for the next 2 Myr. Relating the Hill radius of each debris disk system and the closest distances reached by the two sources, we defined the flyby events in terms of position and time.}
   {We find that in the period between the last 5 Myrs and the next 2 Myrs, 90\% of the analyzed systems have experienced at least a close flyby, while 7\% of them have experienced flybys at distances greater than $~0.5 R_{\rm{\,Hill}}$. In particular, 75\% of them have experienced at least one past close encounter and 36\% multiple past close encounters. From the sub-sample of resolved debris disk (41 out of 94), 80\% of the analyzed systems experience at least an encounter within 0.8 pc. From the subsample of 10 debris disks with planets, half of these systems do show misalignments between disk and planet, stirring,  or asymmetries. Systems with a misalignment between the planetary orbit and the disk do indeed experience at least one flyby event. In particular, when the planet orbits have a difference with the disk inclination higher than about 20$\,^\circ$, as in the case of HD 38529, we find that multiple close encounters have taken place in the last 5 Myr, as theoretically predicted.}
   {The high incidence of encounters, particularly close encounters,  experienced by the systems in the last 5 Myr suggests the fundamental impact of flybys on the evolution of debris disks. Moreover, despite the low statistics, it is interesting to highlight that flybys that have  been theoretically predicted so far in peculiar resolved systems have also been observationally constrained.}


   \maketitle
%

\section{Introduction}
Detected at all ages around very young 1 Myr old stars and up to 10 Gyr, debris disks are usually associated with the last stage of the evolution of protoplanetary disks, where only a remnant of dust remains. This dust is possibly made up of the second generation of material whereby the gas must have already dissipated and a planetary system had already formed. The far- and near-infrared (FIR\ and NIR) excess emission of debris disk systems originates from a cold disk, namely, an analog of the Kuiper Belt in the Solar System, and/or a hot disk, namely, an analog of the zodiacal dust \citep[e.g.,][]{Hughes2018}. This general picture has been changing over the last decade, since the gas has also started to be commonly detected in debris disks after the first detection of CO observed for HD 9672 \citep{Dent2005}. 

A variety of structures have been resolved not only in protoplanetary disks, but also in debris disks \citep[][and references therein]{Hughes2018}. In particular, clumps and asymmetries can be caused either by the presence of companions \citep{Wyatt2007} or, for instance, by flybys \citep{Kalas1996}. 
During a flyby around a protoplanetary disk system, predictions state that there is a possibility that a part of the disk material is captured by the perturber object, leading to the formation of a secondary disk around it \citep[e.g.,][]{Cuello2020, Cuello2021}. 
According to hydrodynamical simulations, the disk behaves differently in terms of the size of the particles and  the pitch angle, depending on whether the encounter is prograde or retrograde \citep{Cuello2020}.
Specifically, prograde encounters lead more frequently to the formation of spirals, while retrograde encounters are expected to be associated with warps. Also, the concentration of dust in the spiral arms could be linked to an enhancement in planetary formation; in the case of the high frequency of close flybys, planetary systems could be then more common. The flyby distance and the mass ratio influence the amount of material that is exchanged between the main star and the perturber. 

In this letter, we aim to  observationally constrain how flybys can be responsible for the formation and evolution of debris disk systems. The high precision of {\it Gaia} eDR3 astrometry and radial velocity of debris disks, presented in Section~\ref{gaiaob}, have been used to reconstruct the trajectories of about 70\% of the source of the catalog with reliable data, together with the trajectory of sources in a certain sky area around the debris disks. In Section~\ref{analysis}, we describe our analysis and results. We summarize our findings in Section~\ref{res}. 

\section{{\it Gaia} eDR3 observations of debris disks}
\label{gaiaob}
We compiled a sample of 254 debris disks from surveys available in the literature at different wavelengths, from the infrared to the millimeter with Spitzer \citep[][]{Meshkat2017, Chen2014}, Herschel \citep[][]{Booth2013, Vican2016, Morales2016}, and ALMA \citep{vanderPlas2019, Marino2020}. All of them show a {\it Gaia} \citep[][]{gaia2016} eDR3 counterpart, while about half of the original sample (118 sources) have reliable astrometry and radial velocity in the {\it Gaia} eDR3 catalog \citep[e.g.,][]{gaia2021}. 

Since the stellar photocenter can generally be altered by the presence of a peculiar activity on the star's surface or close to it, we first investigated the effect of the debris material around the central object in {\it Gaia} astrometric measurements. The spectral energy distribution (SED) of debris disks is characterized by the stellar photosphere and mid- to far-infrared (MIR-FIR) excess due to the debris material around the central star, even though the peak of the {\it Gaia} band between 330-1050 nm covers mostly the photosphere of A to M type stars. This implies that only the contribution from the inner material can affect the SED in {\it Gaia}'s range. In the first approximation, we computed the ratio between the flux density of stars and disks\footnote{
$\int\,F_{\,\rm disk}\,d\lambda$/$\int\,F_{\,\rm star}\,d\lambda$} in the {\it Gaia}  wavelength ranges. Using a simulation of debris disk systems (DMS; \citealp{Kim2018}), we varied the spectral type of the central star and the disk properties (e.g., inner and outer radius of disks consistent with those of the systems in our sample) 
as well as the dust parameters (e.g., blow-out grain size, slope of the grain size distribution, and dust composition). 

We find that this ratio lies between 10$^{-3}$ to 10$^{-5}$, which means that the contribution from debris disks on the SED in the {\it Gaia} bands is negligible. Consequently, we conclude that the motion of the star photocenter cannot be altered by the debris material. We note that this ratio could increase only in the case of a disk more massive with at least two orders of magnitudes than the typical disk mass (e.g., 10$^{-9}$ to several 10$^{-7}$~$M_{\odot}$; \citealp{Greaves2005}, and references therein) or with a high abundance of small grains with a higher slope of the grain size distribution than that expected from a standard equilibrium collisional cascade (i.e., -~3.5; \citealp{Dohnanyi1969}). 

\section{Analysis: Flybys}
\label{analysis}
The {\it Gaia} catalog allows us to reconstruct the flyby events \citep[e.g., ][]{Ma+2022, BailerJones2018}, taking into account a large sample of potential perturbers. In particular, we derived the linear galactocentric coordinates and velocities\footnote{using the Python coordinate conversion \tt{galactocentric}} from the position, parallax, proper motions, and radial velocity from {\it Gaia}. A flyby takes place when the minimum distance between the two sources becomes smaller than the Hill Radius, 
$R_{\rm {\,H,i}}$, 
which is defined as the gravitational sphere of influence of the $i$ star. 

The distance between possible flyby objects and the main star is computed every 45000 years. This means that assuming a realistic relative proper motion around 10 km\,s$^{-1}$, there would be a new calculation for every 0.35 pc of change in distance between the stars. This is less than the measurement of any calculated Hill Radius and also appears to be a good sampling rate for completeness. 

Areas of the sky are defined for all 118 stars bearing debris disks and having a DR2 radial velocity measure. Those belonging to an association are grouped together, while those that do not belong to any association are distinguished into five groups. We adopted the quantitative definition of the Hill Radius from \cite{Ma+2022}:
\begin{equation}\label{eq:7}
R_{\rm{\,H}}\sim \left(R\right) \left(\frac{m_{\rm{\,main}}}{3M_{\rm{\rm{\,enc}}}\left(R\right)}\right)^{\frac{1}{3}},
\end{equation}
where $m_{\rm{\,main}}$ is the mass of the star, $M_{\rm{\rm{\,enc}}}\left(R\right)$ is the part of the mass of the galaxy which is enclosed within the stellar orbit, and $R$ is the distance of the star from the galactic center defined as
\begin{equation}
R=\sqrt{X^2+Y^2+Z^2},
\end{equation}
where X, Y, and Z are the linear galactocentric coordinates of the star derived from the Gaia eDR3 catalog. 
For most of the sources, we consider the mass of the central star compiled in the TESS catalog \citep{TESS}. $M_{\rm{\rm{\,enc}}}\left(R\right)$ is calculated assuming a spherical distribution of the mass of the Milky Way, which leads to the following formula: 
\begin{equation}
M_{\rm{\rm{\,enc}}}\left(R\right)=\frac{R V^2\left(R\right)}{G},
\end{equation}
where $R$
is the distance of the star from the galactic center, $G$ is the gravitational constant, and $V\left(R\right)$ is the circular velocity, which can be approximated as the following linear function: 
\begin{equation}\label{eq:8}
V\left(R\right) = \left(229.0 \pm 0.2\right) - \left(1.7 \pm 0.1\right)\left (R - R_\odot \right),
\end{equation}
 where $R_\odot$ is the distance of the Sun from the galactic center  \citep[as in, e.g.,][]{Eilers2018}. 
\begin{table}
\centering
\renewcommand{\arraystretch}{1.2}
\begin{tabular}{|l|r|r|r|}
\hline
  Name & $M_{\rm{main}}$ [M$_{\odot}$] & $R_{\rm{H}}$ [pc] & $dR_{\rm{H}}$ [pc] \\
\hline
  HD    377 & 1.07 & 1.271 & 0.005\\
  HD  35650 & 0.66 & 1.083 & 0.005\\
  AU Mic & 0.66 & 1.083 & 0.005\\
  $\eta$ Crv & 1.48 & 1.411 & 0.004\\
  BD+49  1280 & 0.597 & 1.05 & 0.02\\
  HD 104860 & 1.11 & 1.292 & 0.009\\
  *q01 Eri & 1.17 & 1.31 & 0.01\\
  HD 107146 & 1.06 & 1.267 & 0.005\\
  HD 112810 & 1.36 & 1.37 & 0.01\\
  HD 128311 & 0.81 & 1.16 & 0.01\\
\hline
 \multicolumn{4}{|c|}{..............................................}\\
\hline\end{tabular}
\caption{\label{RaggiHill} Masses \citep[from ][]{TESS} of the main objects are given in the first column, while the computed Hill Radii and errors in the second and third columns. 
The full table is available online.}
\end{table}

The mass and computed Hill Radius are given in Table~\ref{RaggiHill}. Since possible flybys between 5 Myr in the past and 2 Myr in the future are being considered, the distance of the star from the galactic center changes over this time. This provides an uncertainty on the determination of the Hill radius, which is the only origin assumed for the error given in Table~\ref{RaggiHill}.

The position of the encounter is computed  by subtracting the galactocentric position defined by $X_{\rm{\,i}}$, $Y_{\rm{\,i}}$, and $Z_{\rm{\,i}}$ of each one of the $i$ objects belonging to the area of the sky, from the position of the main star defined by $X_{\rm{\,main}}$, $Y_{\rm{\,main}}$, $Z_{\rm{\,main}}$. 
The new position for the perturber and each debris object are computed, respectively, as: 
$X_{\rm{\,i}}$ + $V_{\rm{\,X, i}}\,t$, $Y_{\rm{\,i}}$ + $V_{\rm{\,Y, i}}\,t$, $Z_{\rm{\,i}}$ + $V_{\rm{\,Z, i}}\,t$, and $X_{\rm{\,main}}$ + $V_{\rm{\,X, main}}\,t$, $Y_{\rm{\,main}}$ + $V_{\rm{\,Y, main}}\,t$, $Z_{\rm{\,main}}$ + $V_{\rm{\,Z, main}}\,t$, 
where $V$ is the velocity and $t$ is the time step of each iteration of $700$ years. 

From the Hill radius of the main object (given in Eq.~\ref{eq:7}), the position and time of the flyby are thus defined as $X_{\rm{\,F,i}}$, $Y_{\rm{\,F,i}}$, $Z_{\rm{\,F,i}}$, and $t_{\rm{\,F,i}}$. The errors in position and time of the encounter are computed from a Monte Carlo simulation, extracting 50 random values of parallax, position, proper motion, and radial velocity within a Gaussian distribution centered at the values and errors from Gaia of parallax, position in the sky, proper motion, and radial velocity (corresponding to the galactocentric position $X_{\rm{\,i}}$, $Y_{\rm{\,i}}$, $Z_{\rm{\,i}}$ and to $V_{\rm{\,X,i}}$, $V_{\rm{\,Y,i}}$, $V_{\rm{\,Z,i}}$). For each of the 50 cases, we calculated the closest distance reached by the perturber during the flyby and the time of the flyby. The lower and upper errors are given  by the difference, respectively, between the minimum and the maximum value found with this method and values of $X_{\rm{\,F,i}}$, $Y_{\rm{\,F,i}}$, $Z_{\rm{\,F,i}}$, $t_{\rm{\,F,i}}$. 
About 80\% of the perturbers have an error lower than 1 pc for the flyby distance and  99\% of the perturbers lower than 1 Myr for the flyby time.

We highlight that the presented methodology considers the multiple flybys as single events that had not been perturbed by any previous encounter. 
The reconstruction of the trajectory of the main star is reliable after the most recent flyby -- and even earlier than that when the last perturber has a mass of $M_{\rm{\,perturber}}<<M_{\rm{\,main}}$. This first approach presented in this letter is thus fundamental in understanding that if at least a close flyby took place during the evolution of a system and its debris disk. Under the assumption of a negligible acceleration, this method defines an upper limit on the time of the flyby, since the effect of acceleration would be only to speed up the star motion, without changing the direction. This method is capable of identifying multiple flybys as single events since it is not reconstructing the deviation of the motion direction that takes place during a flyby. 
Potential perturbers are searched for in a different area of the sky: either in a region of the association where a debris disk belongs to it or, for isolated sources, in different regions of the sky that include the main source; specifically, four regions of 90$^\circ$x90$^\circ$ for sources with parallaxes larger than 7.4 mas, and a fifth region of 80$^\circ$x50$^\circ$ for sources between 3 and 10 mas in parallax (more specific limits in latitude, longitude, and parallax are given in Appendix A). 
A close encounter is defined as an encounter where the minimum distance is $d_{\rm{\,min}}$~<~0.5 $R_{\rm{\,Hill}}$ \citep[e.g.,][]{Ma+2022}.

To validate our methodology, we used the observations of HD 106906 presented by \citet{DeRosa+2019} and \citet{Ma+2022}. Since only the radial velocity of HD 106444 is in Gaia DR2 (and consequently in Gaia eDR3), we take the radial velocity from \citet{Song+2012} and \citet{Gontcharov2006a}, respectively, for HIP 59721 and HD 106906.
We also confirm the two expected flybys for these stars, corresponding to HD 106444 (flyby at 0.19 pc 2.96 Myr back in time, in agreement with the expected flyby at $0.65^{\,+0.62}_{\,-0.90}$ pc and $3.49^{\,+1.75}_{\,-0.90}$ Myr back in time), and to HIP 59721 (flyby at 0.66 pc and 1.55 Myr back in time, while \citet{Ma+2022} found a minimum distance of 0.71 + 0.18 pc at 2.19 - 1.03 Myr back in time). \\

\section{Results}
\label{res}
From the linear velocities in 3D of the main objects and perturbers, the relative velocities were found to be in the range of 10-100 km\,s$^{\rm -1}$. 
Taking into account all the flybys within 5 Myrs and in the next 2 Myrs, 90\% of the analyzed systems have at least a close flyby (as defined in Section~\ref{analysis}), while 7\% are flybys at distances higher than $~0.5\,R_{\rm{\,Hill}}$. In particular, if we concentrate on close encounters, we find that 75\% of the debris disk systems experienced at least one close flyby, while 62\% are expected in the near future; moreover, 45\% of the systems experienced both past and near-future close encounters.  The histograms in  Fig. \ref{Histoflybys} show that past flybys are taking place at any age between $\sim$1 Myr and  $\sim$5 Gyrs and around stars with all spectral types between F0 and M1. Moreover, 39\% has one close encounter in the past, as seen from Fig. \ref{Histoflybys}, while  36\% has multiple past close encounters.

\begin{figure*}[htbp]
\centerline{\includegraphics[height=24cm]{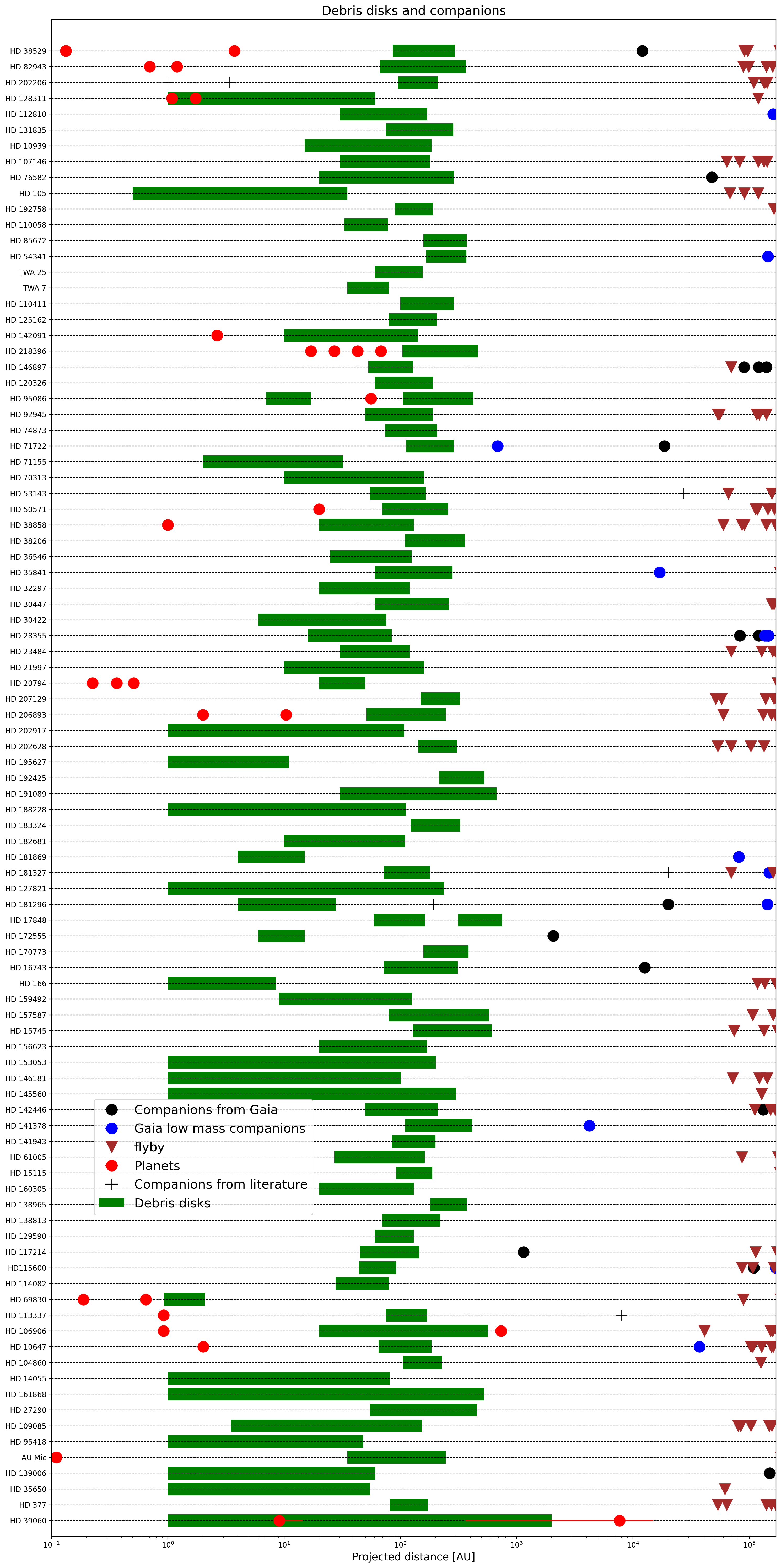}}
\caption{Sizes of the resolved debris disk systems and the positions of planets, companions, and flybys from the central object. On the x-axis, the projected distance from the main star is given in AU. In the y-axis, all systems with resolved debris disks are listed. The panel overlaid in the figure explains the different symbols and colors. }
\label{Debrisgraphflyby}
\end{figure*}

For the 94 systems with resolved debris disks, we can proceed with a more detailed analysis.  Figure~\ref{Debrisgraphflyby} shows the size of the disks, positions of planets, and low ($M$ $<$ 0.4 $M_{\odot}$) and high-mass companions compiled from the literature. The distances of the flybys of the systems with radial velocities (41 out of 94) which happened in the last 5 Myr are also shown. 
We find that 80\% of the analyzed systems experienced at least an encounter.\\ 
Compiling the 
spectral energy distributions (SEDs) of the 162 perturbers of all the systems that caused close encounters within the last 5 Myr, we find that 25\% of all of them do show a FIR excess over the stellar photosphere. Since the orbital change of planetesimal (e.g., eccentricity and inclination) can result from the existence of nearby sculpting planet through close encounters (\citealp{Larwood2001, Kalas2007, Dong2020}), possibly leading to the change of disk's vertical scale height (\citealp{Quillen2007}); this could be a signature of debris material around the central star.

Moreover, we consider the ten debris disk systems with known planets: \object{HD 38529}, \object{HD 38858}, \object{HD 50571}, \object{e Eri}, \object{*q01 Eri}, \object{HD 82943}, \object{HD 69830}, \object{HD 128311}, \object{HD 206893}, and \object{AU Mic}. Half of these systems do show misalignments between disk and planet \citep{Xuan2020}, stirring \citep{Moor2015}, or asymmetries \citep{Lovell2021, Marino2020} in the disks. These have been theoretically interpreted by self-stirring, past flybys \citep{Moor2015, Cuello2020}, or companions. We can now observationally check the frequency of past flybys in these systems. 

\subsection{Single debris disks hosting planets}
In the system of $\boldmath{HD\,38529}$ there is a difference in inclination between the disk and planetary system of at least $\Delta_{\rm{\,i}}$ = 21 - 45$^\circ$ (\cite{Xuan2020}). We note that we found several possible flybys for this star, with the closest one being caused by Gaia eDR3 6530376419268321280, reaching a minimum distance of $d_{\rm{\,min}}=0.44^{\,+2.5}_{\,-0.4}$ pc $2.97^{\,+0.11}_{\,-0.19}$ Myr back in time. This means that there is a chance that the object could have interacted with the disk and the planets. This object has an absolute magnitude of $G = 8.788 \pm 0.003$ mag. Other possible past flybys we found are the ones caused by Gaia eDR3 3241155739457638272 $0.54 \pm 0.01$ Myr back in time at a distance of $d_{\rm{\,min}}=0.51^{\,+0.14}_{\,-0.13}$ pc, by PMJ05428+1613E $0.18\pm0.01$ Myr back in time at a distance $d_{\rm{\,min}}=0.88^{\,+0.26}_{\,-0.08}$ pc, and by HD 246651 $0.37\pm0.01$ Myr back in time at a distance of $d_{\rm{\,min}}=0.96^{\,+0.51}_{\,-0.32}$ pc. The companion, HD 38529B, which lies at $d=12018 \pm 21$ AU in projected separation and could be older than the main star according to our isochrones, does not have a  Gaia DR2 radial velocity measurement. According to the analysis from \cite{Xuan2020}, interactions between the planets, disk, and companion could explain the misalignment, either through disk warping or through a misalignment between the planes of the companion and of the outer planet, ultimately leading to the misalignment of the disk. 

Instead, $\boldmath{HD\,82943}$ shows coplanarity between the planets and the disk. In fact, the inclination of the orbit of the planets $I_{\rm{\,planets}}=20\pm4$ deg is in line with the inclination of the disk $I_{\rm{\,disk}}=27\pm4$ deg (\citealp{Kennedy2013}). We find possible close flybys caused by CD-26 8683 $0.29 \pm 0.01$ Myr back in time at a distance of $d_{\rm{\,min}}=0.51^{\,+0.17}_{\,-0.3}$ pc and by BD-02 2645 $0.24 \pm 0.01$ Myr back in time at a distance of $d_{\rm{\,min}}=0.47^{\,+0.01}_{\,-0.03}$ pc. Then we have 
$\boldmath{HD\,128311}$ which hosts two giant planets, with masses around 1-3 $M_{\rm{\,J}}$, with the orbits in a 2:1 resonance \citep[][]{MoroMartin2010}. The inner radius of the disk is external to the orbit of the planets. We find a possible flyby caused by Wolf 295 A at a distance $d_{\rm{\,min}}=0.29^{\,+0.17}_{\,-0.14}$ pc $1.04^{\,+0.02}_{\,-0.01}$ Myr back in time. $\boldmath{HD\,50571}$ hosts a disk resolved by Herschel at 70, 100, and 160~$\mu$m. The nature of this disk has been explained by \citet{Moor2015} by self-stirring of planetesimals in the early stages of the protoplanetary disks: the larger ones influence the trajectories of the smaller objects leading to a collisional cascade. Since the growth time of the planetesimals is proportional to $\frac{P}{S}$, with $P$ being the period and $S$ the surface density of the disk, the cascade begins in the inner disk and propagates outwards \citep{Moor2015}. Companions and stellar flybys could be the reason for additional stirring linked to other collisional cascades, but we did not find any companion. The planet, undetected by Gaia, is a $30 \pm 15$ $M_{\rm{\,J}}$ brown dwarf orbiting at $d=20.3\pm0.3$ AU \citep{Konopacky2016}. The closest flyby happens at a distance $r_{\rm{\,min}}=0.55^{\,+0.34}_{\,-0.21}$ pc $0.68\pm0.01$ Myr back in time with the transit of iot Hyi, with another close flyby being caused by TYC 8854-1126-1 further away back in time ($1.79^{\,+0.04}_{\,-0.07}$ Myr ago at a distance of $d_{\rm{\,min}}=0.58^{\,+0.46}_{\,-0.17}$ pc). CD-60 6470A is another possible source for a flyby, with an encounter $0.59^{\,+0.03}_{\,-0.02}$ Myr back in time at a distance $d_{\rm{\,min}}=1.29^{\,+2.76}_{\,-1.13}$ pc. We also find that $\boldmath{e Eri}$ hosts a very faint disk \citep{Kennedy2015} and at least three (possibly six) inner super-Earths \citep{Feng2017}. We do not find any close encounters in the past 5 million years for this object, nor do we find any companions either.

$\boldmath{HD\,206893}$ had possible close encounters with HD 190617 ($d_{\rm{\,min}}=0.30^{\,+0.13}_{\,-0.03}$ pc $0.31\pm0.01$ Myr back in time), HD 189242 ($d_{\rm{\,min}}=0.46^{\,+0.48}_{\,-0.17}$ pc $1.24\pm0.01$ Myr back in time) and HD 182261 ($d_{\rm{\,min}}=0.65^{\,+3.55}_{\,-0.5}$ pc $2.18^{\,+0.34}_{\,-0.23}$ Myr back in time) from our analysis. The system has a brown dwarf at 11 AU and another one at 2 AU; the orbit of the outer one is likely eccentric, with $e=0.14^{\,+0.05}_{\,-0.04}$ \citep{Marino2020}. The disk has a gap at 74 AU, which could be explained by the presence of a Jupiter-like planet, and asymmetries with a brighter NE arm which could be explained by the companion \citep{Marino2020}. The impact of possible flybys hasn't been studied in the literature. $\boldmath{HD\,69830}$ experienced an encounter with HD 57901 at $d_{\rm{\,min}}=0.43^{\,+0.06}_{\,-0.02}$ pc at $0.31\pm0.01$ Myr back in time. The system has known planets and an asteroid belt within 3 AU from the star. We find that $\boldmath{*q01\,Eri}$ shows a flyby with HD 23356 at $d_{\rm{\,min}}=0.5^{\,+0.09}_{\,-0.07}$ pc, $0.31 \pm 0.01$ Myr back in time and another one with HD 39855 $0.61\pm0.01$ Myr ago at a radius $d_{\rm{\,min}}=0.52^{\,+0.13}_{\,-0.08}$ pc. A 0.2 - 0.4~$M_{\rm{s}}$ companion, WT 50, is found at $37350\pm29$ AU in projected separation. Asymmetries in the disk could be modeled by an eccentric disk, the presence of a clump on the inner SW edge, or a combination of both. The clump shows the most consistency with the data and could have originated from a disk-planet interaction or from a collision \citep{Lovell2021}. Since the asymmetry is in the inner part of the disk, a planet-companion seems a much more likely explanation than a flyby.

We note that $\boldmath{AU Mic}$ was not shown to experience any close past or future encounters, but only multiple flybys at distances greater than 0.5 $R_{\rm Hill}$. More precisely, in the last 5 Myr, the most recent flyby was caused by Gaia EDR3 6427708902553822592 $0.17\pm0.01$ Myr ago at a radius of $d_{\rm{\,min}}=0.93^{\,+0.06}_{\,-0.05}$ pc, along with two other flybys by HD 283750 $0.7^{\,+0.2}_{\,-0.1}$ Myr ago at a radius of $d_{\rm{\,min}}=0.90^{\,+2.0}_{\,-0.03}$ pc and  Gaia EDR3 436648129323934592 $0.28\pm0.01$ Myr ago at a radius of $d_{\rm{\,min}}=0.96^{\,+0.11}_{\,-0.05}$ pc, respectively. This system is a young M star (20-30 Myr old) and clumps have been observed in it, as well as a Neptunian planet detected with TESS \citep[][]{Plavchan+2020}. A second planet, AU Mic c, with an orbital period of 18.86 days has been recently discovered  \citep{Gilbert+2022, Martioli+2021} and confirmed \citep{Wittrock+2022}. A new planet, AU Mic d, has been discovered and it is now undergoing validation \citep{Wittrock+2023}.

\section{Discussion and conclusions}
The main result of our study concerns the extremely high fraction of flyby events experienced by debris disks: in the last 5 Myrs and in the next 2 Myrs, we find that 90\% of the analyzed systems have experienced at least one close flyby, while 7\% flybys have taken place at distances higher than $~0.5 R_{\rm{\,Hill}}$. 
Moreover, 80\% of the subsample of resolved debris disks experienced at least a close flyby event (i.e., within 0.8 pc) in the last 5 Myrs. This especially high frequency is not preferentially found at an age or spectral type. It is important to highlight that in our approach, we consider possible perturbers of the main debris disk to be all  spatially selected sources -- and not only cluster members, as done, for instance, by \citet{Ma+2022}. 
We note that this approach might be valid in the first approximation for an isolated cluster. This situation has been also simulated to constrain the effect of  flybys in the cluster dynamics and protoplanetary disk evolution. In particular, in the case of low-mass clusters \citet{Pfalzner+2021} found a higher frequency of  close flybys compared to previous studies.  
On the one hand, it will be now important to reconstruct in 3D the dynamical history via 3D flybys in the young systems at different ages and different evolutionary phases of the disks in order to understand whether this is a fundamental effect that needs to be taken into account. On the other hand, when concentrating only on clusters, it will be fundamental to also reconstruct  the dynamical history of sources without disks today to explore the impact of destructive flyby events in young clusters.

Moreover, from the spectral energy distribution of all the 162 perturbers, we find that 25\% of them that have experienced a close encounter within the last 5 Myr show a FIR excess over the stellar photosphere. This might be the first signature of debris material around the perturbers which may originate from the exchange of dust between the main star with its debris disk and the perturber. This dust exchange has been theoretically predicted by 3D simulations, which suggest a lower dispersion of dust for the primary disk when the perturber already has a secondary disk \citep[][]{Picogna+2014}. The full 3D geometry is fundamental to understand mass loss. From these simulations, it can be seen that in the case of two similar disks, up to 15\% of the mass of the disk can be exchanged, with a dependence on the angle of incidence \citep[][]{Picogna+2014}. Studying debris disks in binary systems,  \citet{Thebaultetal2021} also found that a dust exchange from the primary to the secondary disk is responsible for a circumsecondary ring.

The subsample of ten debris disks with known planets and resolved disks forms a basis for the  discussion of the influence of binarity, planet formation, and flybys in the evolution of debris disks. 
The misalignments between the planes can be caused by interactions with stellar companions or other planets, which could excite eccentricities and inclinations for the orbits of the planets and of the disk. As discussed in \citet{Picogna+2014}, misalignments are considered to be a particular signature of a past flyby, with a maximum value of tilting of 9$^\circ$ between the orbits of the planet and the disk. Exceptionally, they could be even larger in the case of multiple past flybys. A primordial origin of such misalignments has been discussed (e.g., \citealp{Kennedy2013}). Such a misalignment between the plane of the planets and the disk can also be the only remaining signature of the flyby for the orbit of the planets. Simulations also show that the encounter could temporarily cancel gaps created by the planets or by companions, as other preexisting structures \citep[][]{Picogna+2014}.

Warping and puffiness in the disk shape could be caused by an inclined companion on a very small timescale, namely, lower than 5 Myr. For example, in the case of HD106906, \citet{Moore+2022} found a recent close encounter with a free-floating planet to be responsible for a companion with orbital parameters. This is in agreement with observations of HD106906b, as we found in our observational constraints. 
\begin{appendix}
\section{Additional material}
\begin{figure*}
        \includegraphics[width=9cm]{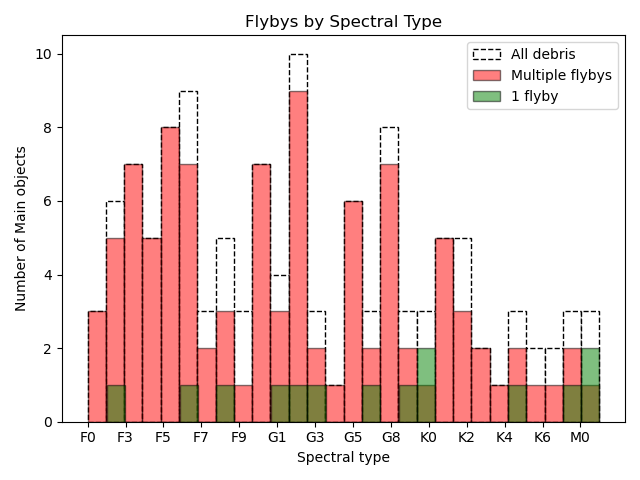}
        \includegraphics[width=9cm]{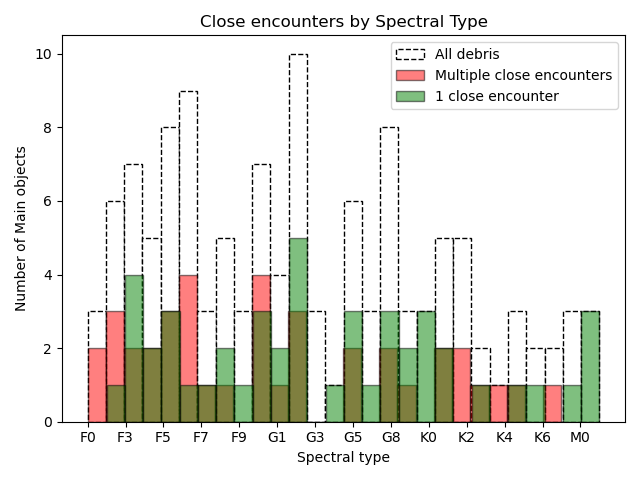}
        \includegraphics[width=9cm]{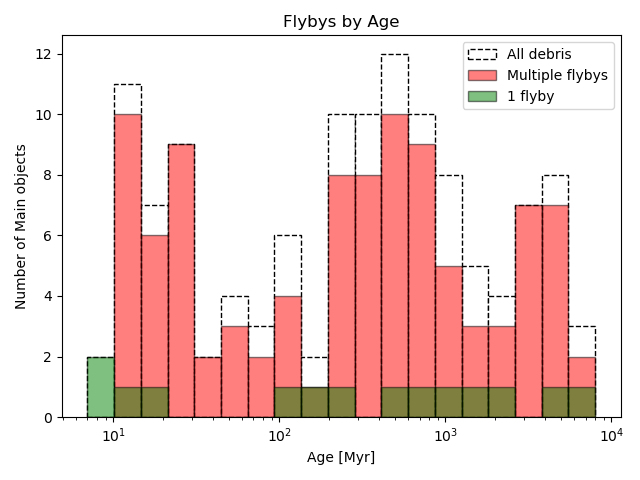}
        \includegraphics[width=9cm]{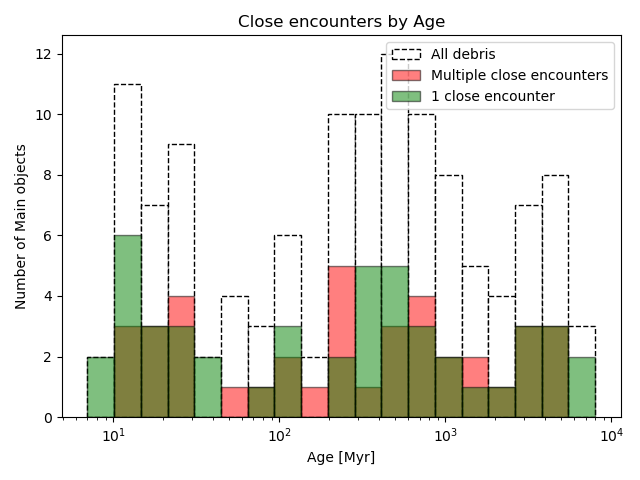}
\caption{Distribution of ages and spectral types of flybys and close encounters in the last 5 Myrs.}
\label{Histoflybys}
\end{figure*}

In this section, we show some additional material complementary to the results discussed in the main body of this letter. In particular, the histograms of single and multiple flybys as a function of the spectral type and age of the main source are shown in Figure \ref{Histoflybys}, highlighting the close encounters. Table~\ref{list_FB} shows the first lines of a complete table given online, where we show the results of all the perturbers responsible for a flyby for each debris disk system. 


\begin{table*}
\renewcommand{\arraystretch}{1.2}
\begin{tabular}{|r|c|c|c|c|c|}
\hline
  \multicolumn{6}{|c|}{Lower Centaurus Crux} \\
\hline
  \multicolumn{1}{|r|}{Main object} &
  \multicolumn{1}{c|}{flyby object} &
  \multicolumn{1}{c|}{$R{_{\rm{min}}}$ [pc]} &
  \multicolumn{1}{c|}{$dR{_{\rm{min}}}$ [pc]} &
  \multicolumn{1}{c|}{$t{_{\rm{flyby}}}$ [Myr]} &
  \multicolumn{1}{c|}{$dt{_{\rm{flyby}}}$ [Myr]} \\
\hline
  HD 115600 * & UCAC4 154-122270 & 0.24 & +0.58-0.03 & -0.44 & +0.08-0.08\\
  * & Gaia EDR3 5868438533566601088 & 0.3 & +1.39-0.11 & 0.55 & +0.13-0.13\\
   & Gaia EDR3 5869330679822111744 & 1.12 & +4.11-0.38 & 1.19 & +0.41-1.39\\
  * & Gaia EDR3 6061962578725632640 & 0.52 & +0.19-0.05 & 0.3 & +0.22-1.86\\
   & TYC 8674-2317-1 & 0.96 & +0.36-0.03 & 0.33 & +2.87-0.18\\
   & SCR J1342-5714 & 0.99 & +0.4-0.22 & -0.18 & +0.01-0.01\\
   & HD 114835 & 0.75 & +0.12-0.08 & 0.31 & +0.04-0.04\\
   & HD 104309 & 1.16 & +1.33-0.05 & 1.58 & +0.01-0.01\\
  HD 114082 & Gaia EDR3 6108901825287594368 & 1.32 & +1.25-0.17 & -1.05 & +0.04-0.06\\
   & CD-57  5422 & 1.06 & +7.58-0.07 & -1.92 & +0.69-0.08\\
   & HD 118771 & 1.21 & +0.37-0.2 & -0.18 & +0.01-0.01\\
   & HD 120056 & 1.05 & +0.37-0.18 & -0.3 & +0.01-0.01\\
   & HD 311959 & 1.33 & +0.27-0.25 & 0.38 & +0.03-0.04\\
  HD 112810 & HD 112794 & 0.76 & +0.09-0.07 & -0.39 & +0.09-0.09\\
   & CD-47  8072 & 1.14 & +0.45-0.51 & 0.33 & +0.02-0.01\\
   & UCAC4 219-066881 & 1.28 & +0.64-0.78 & 0.73 & +0.02-0.04\\
  HD 117214 * & Gaia EDR3 6080914894977456768 & 0.55 & +3.12-0.21 & 0.98 & +0.07-0.08\\
   & Gaia EDR3 6063534330582947328 & 0.77 & +0.1-0.12 & 0.17 & +0.01-0.01\\
  * & SCR J1342-5714 & 0.43 & +0.23-0.03 & -0.09 & +0.01-0.01\\
   & HD 103974 & 0.84 & +0.63-0.49 & 1.79 & +0.01-0.01\\
\hline
\multicolumn{6}{|c|}{..............................} \\
\hline\end{tabular}
\centering
\caption{\label{list_FB} 
Distance and time of the flybys found for each  debris disk system highlighting whether they belong to an association or to with part of the sky considered. Close encounters correspond to the [*] points in the first column. Names in bold are objects with FIR excess. The full table is available at the CDS.}
\end{table*}

\end{appendix}
\begin{acknowledgements}
This work has made use of data from the European Space Agency (ESA) mission
{\it Gaia} (\url{https://www.cosmos.esa.int/gaia}), processed by the {\it Gaia}
Data Processing and Analysis Consortium (DPAC,
\url{https://www.cosmos.esa.int/web/gaia/dpac/consortium}). Funding for the DPAC
has been provided by national institutions, in particular, the institutions
participating in the {\it Gaia} Multilateral Agreement. 
The authors thank our referee Giovanni Picogna for the comments and suggestions that helped to improve this letter. MK gratefully acknowledges the funding from the Royal Society. 
\end{acknowledgements}

%
   \bibliographystyle{aa} 
   \bibliography{dd-flybys_bib} 
%

\end{document}